\documentclass[prl,twocolumn,amsmath,amssymb,superscriptaddress]{revtex4-2}

\usepackage{graphicx}
\usepackage{dcolumn}
\usepackage{bm}
\usepackage{mathrsfs}
\usepackage{natbib}
\usepackage{amsmath}
\usepackage{amssymb}
\usepackage{Jonasmacros}
\usepackage{mathptmx}
\usepackage{txfonts}
\usepackage{ifsym}
\usepackage[usenames,dvipsnames,svgnames]{xcolor}

\usepackage[colorlinks,citecolor=magenta,linkcolor=red]{hyperref}
\usepackage[usenames,dvipsnames,svgnames]{xcolor}

\renewcommand{\section}[1]{\emph{#1}:}

\begin{document}
\title{Chiral Phonons Enhance Ferromagnetism}

\author{Jonas Fransson}
\affiliation{Department of Physics and Astronomy, Uppsala University, Box 516, 751 21 Uppsala, Sweden}

\author{Yael Kapon}
\author{Lilach Brann}
\author{Shira Yochelis}
\affiliation{Department of Applied Physics, The Hebrew University, Jerusalem 9190401, Israel}

\author{Dimitar D. Sasselov}
\affiliation{Harvard-Smithsonian Center for Astrophysics, Cambridge, MA 02138, USA}

\author{Yossi Paltiel}
\affiliation{Department of Applied Physics, The Hebrew University, Jerusalem 9190401, Israel}

\author{S. Furkan Ozturk}
\affiliation{Harvard-Smithsonian Center for Astrophysics, Cambridge, MA 02138, USA}
\affiliation{King’s College, Cambridge, CB2 1ST, UK}

\begin{abstract}
Recent experiments suggest that the conditions for ferromagnetic order in, e.g., magnetite, can be modified by adsorption of chiral molecules. Especially, the coercivity of magnetite was increased by nearly 100 \%, or 20 times the earth magnetic flux density, at room temperature. The coercivity was, moreover, demonstrated to increase linearly with temperature in a finite range around room temperature. Based on these results, a mechanism is proposed for providing the necessary enhancement of the magnetic anisotropy. It is shown that nuclear vibrations (phonons) coupled to ferromagnetic spin excitations (magnons) absorb the thermal energy in the system, thereby diverting the excess energy that otherwise would excite magnons in the ferromagnet. This energy diversion, not only restores the ferromagnetic order but also enhances its stability by increasing the anisotropy energy for magnon excitations. The coupling between phonons with magnons is enabled by chirality due to the lack of inversion symmetry.
\end{abstract}



\maketitle


Ferromagnetic order typically form below a critical temperature, $T_c$, above which higher energy spin excitations become occupied due to thermal fluctuations. Such spin excitations, often referred to as magnons, tend randomize the order, hence, having a detrimental effect on the magnetic state. As the amount of evidence for this behavior that has been collected over the years is overwhelming, the perception as ferromagnetism as a low temperature collective phase is convincing, albeit that $T_c$ may be larger than 1,000 K in some compounds \cite{Keffer1966,JPhysD.49.095001}.

Recently, it was reported that supramolecular aggregates may on the one hand show stable room temperature ferromagnetism and one the other hand small, if not vanishing, magnetic moment at low \cite{JPhysChemLett.7.4988,ACSNano.14.16624}. These truly surprising results were based on a coercive field which increases with increasing temperature. By invoking coupling between localized spin moments and phonons, these results were explained under the condition of broken inversion symmetry such that anharmonic contributions to the phonon excitations become non-negligible \cite{JPhysChemLett.14.2558}. In fact, breaking inversion symmetry opens for non-vanishing spin-dependent electron-phonon interactions \cite{PhysRevResearch.5.L022039,JChemPhys.159.084115} \cite{JPhysChemLett.15.6370}

The phenomenology of magnetism pertains to seemingly separate questions, such as room-temperature ferromagnetism \cite{NatPhysics.5.840,NatNanotech.13.289,JACS.140.11519,APL.92.082508,APL.81.4212,NanoLett.9.220}, topological matter \cite{Science.329.61,Nature.546.270,NatMaterials.19.484,Nature.603.41}, and exotic spin excitations (e.g., skyrmions) \cite{Nature.442.797,Science.323.915,Nature.465.901,ChemRev.121.2857}. Studies suggest that magnetic phenomena are not only pertinent in biological contexts but are also crucial for, e.g., oxygen redox reactions, on which aerobic life depends \cite{PNAS.2202650119, Gupta2022.11.29.518334}. Recently, magnetic phenomena have also been linked to life’s biomolecular homochirality, where the chiral molecular symmetry is broken by processes controlled by electron spin \cite{ozturk2022, ozturk2023crystallization}.

Nuclear vibrations, phonons, are conventionally considered as a source for decohering and dissipating magnetic states and order. It is, therefore, easy to make the connection between increased temperatures and phonon activation which, in turn, leads to energy level broadening, occupation of multiple electronic states with competing magnetic properties, ultimately caused by distortions and reformations of nuclear configurations. This, faulty conclusion is based on the conjecture that there is an intimate relationship between magnons and phonons. As will be shown in this article, such a conjecture is not founded on physical grounds, but merely presumptions. While nuclear vibrations in a previous publication has been considered as a source for stabilizing magnetic order \cite{JPhysChemLett.14.2558}, it has previously never been demonstrated that nuclear vibrations provide a sink for the energy that would otherwise induce magnon excitations.

The question to be addressed here is whether nuclear (ionic, molecular) vibrations can increase the magnetic stability in structures that are already ferromagnetically ordered. The investigation is stimulated by the experimental observations on ferromagnetic structures that acquires an increasing coercivity with increasing temperature \cite{ACSApplMaterInterfaces.13.34962, JACS.144.7302, kapon2024}.

\begin{figure}[t]
\begin{center}
\includegraphics[width=\columnwidth]{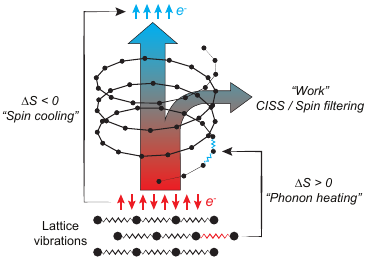}
\end{center}
\caption{We present an illustration depicting the coupling between chiral phonons and electron spin within a thermodynamic framework. Consider an isolated system of chiral molecules on a ferromagnetic surface, where angular momentum-carrying degrees of freedom are distributed. In a microcanonical ensemble, the system’s total energy remains constant; thus, any work performed by the system must be balanced by heat supplied to it. Through chiral-induced spin selectivity (CISS), chiral molecules filter electron spins, reducing the system’s entropy. To maintain energy and angular momentum balance, other degrees of freedom must compensate, with chiral lattice vibrations, phonons, emerging as energetically favorable contributors. Consequently, heating the system to enhance the availability of chiral phonons facilitates the operation of CISS, where the effective cooling by spin-filtering is counterbalanced by effective heating from chiral phonons.}
\label{fig-Schematic}
\end{figure}

In our recent experimental study, we adsorbed chiral molecules onto a ferromagnetic nickel substrate and observed an enhancement in the coercive field of the substrate as the temperature increased \cite{kapon2024}. We provided an explanation for this phenomenon using a thermodynamic model, which we named the \emph{chiral heat engine}. How can we reconcile the \emph{chiral heat engine} model, discussed in Kapon et al. \cite{kapon2024}, with the microscopic theory presented here, which involves lattice vibrations?

The chiral induced spin selectivity (CISS) effect arises from the spin-filtering effects of chiral molecules, which reduce entropy by limiting the available microstates in the electronic spin distribution. In the microcanonical ensemble, the relevant system can be thought of as the angular momentum distribution, encompassing electronic spin and other angular momentum-bearing degrees of freedom. According to the first law of thermodynamics, the internal energy of an isolated system remains constant, so any work performed by the system must be compensated by heat absorption, as seen in Fig. \ref{fig-Schematic}. This thermodynamic framework explains the work-heat relationship, but the role of chiral phonons becomes evident when we consider the nature of the heat input. For angular momentum conservation to hold, the heat entering the system must also carry angular momentum. These heat carriers are not merely phonons but specifically chiral phonons that carry angular momentum.

But how do chiral phonons contribute to CISS? In a semiclassical picture, as an electron propagates along a helical path induced by the chiral potential of a closed-shell chiral organic molecule, it experiences an effective Lorentz force. This force redirects the electronic angular momentum, necessitating a corresponding transfer to maintain angular momentum conservation. Closed-shell chiral organic molecules, being non-metallic, lack delocalized, lower-energy electrons which can accommodate this transfer. Thus, the most accessible repository for angular momentum exchange lies in low-energy chiral lattice vibrations. This dynamic interaction between the electron and the chiral phonons ensures the efficient exchange of angular momentum in a closed system, enabling the electron to maintain its helical trajectory. Thus, chiral phonons play a critical role in facilitating angular momentum conservation while the electronic spin distribution is filtered, effectively fueling the “chiral engine.”

The model presented in this article is aimed at capturing the ferromagnetic properties of the surface magnetism of, e.g., magnetite (Fe$_3$O$_4$) or Ni, onto which a layer of chiral molecules are adsorbed, as experimentally investigated by Ozturk et al. and Kapon et al. \cite{ozturk2023magnetization, kapon2024}. Specifically, the increased coercive field that results from the presence of the molecules, as well as its further increase with the temperature is the target here. Pertaining to magnetite and other local moments ferromagnets, the model is based on localized spin moments, as well as nuclear motion, where both quantities embedded in an electronic structure which constitutes the spin-spin and spin-vibration couplings, discussed in detail in \cite{PhysRevMaterials.1.074404}. As a model for a ferromagnetic lattice, such model pertains very well to magnetite due to the local moment structure. Its relevance to Ni crystals is reasonable despite the lack of as well defined localized moments in Ni crystals in comparison with magnetite.

In compounds like magnetite, the ferromagnetism is a consequence of a long range order among the localized spin moments carried by Fe, where the presence of oxygen provides the ferromagnetic exchange interaction through double exchange. Local moment ferromagnets can be effectively modelled using the anisotropic Heisenberg model $\Hamil_\text{FM}=-\sum_{mn}(J_{mn}\bfM_m\cdot\bfM_n+I_{mn}M_m^zM_n^z)$, where $\bfM_m$ defines the magnetic moment associated with the spatial coordinate $\bfr_m$, subject to isotropic, $J_{mn}$, and anisotropic, $I_{mn}$, interactions with the magnetic moment $\bfM_n$ at $\bfr_n$.

\begin{figure}[t]
\begin{center}
\includegraphics[width=\columnwidth]{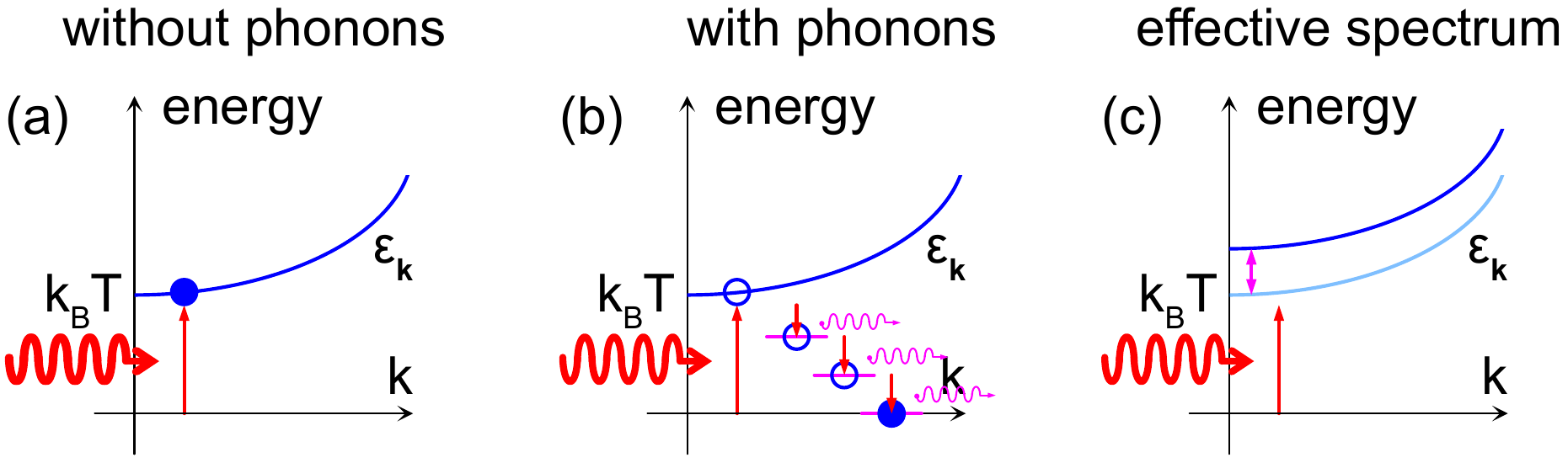}
\end{center}
\caption{Magnon spectrum (a) without and (b) with coupling to chiral phonons, and (c) effective magnon spectrum caused by the magnon-phonon interactions. Without magnon-phonon coupling (a), the thermal energy from the environment, red wiggle, leads to magnon excitations, blue circle. With magnon-phonon coupling (b), the magnon excitation repeatedly loses energy to the phonons, magenta wiggles, successively decaying back to the ground state by occupying intermediate energy levels. Effectively, (c) the phonon induced decay leads to an increased anisotropy.}
\label{fig-MagnonPhonon}
\end{figure}

For two-dimensional structures, pertaining to the present context, this model enables an out-of-plane ferromagnetic ($J_{mn}>0$) ground state as long as the anisotropic interaction $I_{mn}$ is negative. The anisotropic interaction provides an energy gap in the excitation spectrum, Fig. \ref{fig-MagnonPhonon} (a), which may be overcome by, for instance, thermal supply of energy.

Here, we consider a novel mechanism that provides an additional source for anisotropy to the magnetic structure. Quite surprisingly, this mechanism is connected to lattice vibrations, phonons, Fig. \ref{fig-MagnonPhonon} (b), which absorb the thermally supplied energy such that the magnetic excitations remain unoccupied and, thereby, stabilizes the ordered state. Effectively, the magnon-phonon interactions lead to an increased magnetic anisotropy, Fig. \ref{fig-MagnonPhonon} (c). A requirement for phonons to couple to the spin degrees of freedom is broken inversion symmetry \cite{PhysRevResearch.5.L022039}. In the present context, the broken inversion symmetry is granted by a layer of chiral molecules adsorbed on the surface of the ferromagnet, thereby, introducing a coupling between the phonons and local spin moments. Other mechanisms for couplings between spins and the lattice motion have been considered in, for instance, Ref. \cite{PhysRevMaterials.1.074404}.

The layer of chiral molecules is assumed to provide phonons through the normal vibrational modes $\bfQ_m=l_m\bfepsilon_mQ_m$, of the nuclear displacement. Here, the operator $Q_m=b_m+b^\dagger_m$ represents the quantized nuclear displacement, where $b_m$ ($b^\dagger_m$) is the phonon annihilation (creation) operator, whereas $l_m$ defines a length scale and $\bfepsilon_m$ denotes the phonon polarization vector. The phonon background is modelled by the harmonic chiral phonons $\Hamil_\text{ph}=\sum_m\omega_mb^\dagger_mb_m$, where $\omega_m$ is associated vibrational energy. The spin and displacement variables are coupled through the electronic structure in the contribution $\sum_{mn}\bfM_m\cdot\mathbb{A}_{mn}\cdot\bfQ_n$, where $\mathbb{A}_{mn}$ defines a coupling tensor between the magnetic and mechanical degrees of freedom. Making use of the notation for the nuclear displacement, we set $\bfA_{mn}=\mathbb{A}_{mn}\cdot\bfepsilon_nl_n$. Then, the model can be summarized as
\begin{align}
\Hamil=&
	\Hamil_\text{ph}
	+
	\Hamil_\text{FM}
	+
	\sum_{mn}\bfM_m\cdot\bfA_{mn}Q_n
	.
\label{eq-model}
\end{align}

\section*{Results and Discussion}
\subsection*{Magnons}
Deviations from the ferromagnetic ground state can be effectively studied through the magnon spectrum. It is particularly interesting to consider the magnon dispersion and whether the interactions with the phonons can introduce an anisotropy that adds to the immanent anisotropy $I_0$. The reason is that this would give an answer to whether the phonons may increase the coercivity of the magnetic state. To this end, the Holstein-Primakoff transformation is applied, with the longitudinal magnetic moment $M^z_m=M-n_m$ is represented in terms of the deviations $n_m=a^\dagger_m a_m$ from the ground state moment $M$. In this scheme, the ladder operators for increasing and decreasing the magnetic moment is defined by $M_m^+=\sqrt{2M(1-n_m/2M)}a_m$, and $M_m^-=a_m^\dagger\sqrt{2M(1-n_m/2M)}$, respectively, in terms of the magnon annihilation ($a_m$) and creation operators $a^\dagger_m$.

By expanding the model $\Hamil$ in the magnon operators, an approximate magnon-phonon model, can be established as
\begin{align}
\widetilde\Hamil=&
	\Hamil_\text{ph}
	+
	M\sum_{mn}
		\biggl[
			\Bigl(J_{mn}+I_{mn}\Bigr)\Bigl(n_m+n_n\Bigr)
			-
			J_{mn}\Bigl(a^\dagger_ma_n+a^\dagger_na_m\Bigr)
\nonumber\\&
	+
		MA_{mn}^zn_m
		-
		\sqrt{\frac{M}{2}}\Bigl(A_{mn}^+a_m^\dagger+A_{mn}^-a_m\Bigr)
		Q_n
	\biggr]
	.
\end{align}
This expression is obtained by retaining magnon operators up to quadratic order and introducing the notation $A_{mn}^\pm=A^x_{mn}\pm iA^y_{mn}$.

The interaction parameters $J_{mn}$, $I_{mn}$, and $\mathbb{A}_{mn}$ are assumed to all be distance dependent on the form, e.g., $J_{mn}=J(\bfr_m-\bfr_n)$. This enables the plane wave expansion $J_{mn}=-\sum_\bfk J_\bfk e^{-i\bfk\cdot(\bfr_m-\bfr_n)}/N$, and analogously for $I_{mn}$ and $\bfA_{mn}$. Expanding the magnon operators in corresponding plane wave operators, that is, $a_m=\sum_\bfk a_\bfk e^{i\bfk\cdot\bfr_m}/\sqrt{N}$, the model can be written
\begin{align}
\widetilde\Hamil=&
	\sum_\bfk
		\Bigl(
			\dote{\bfk}a^\dagger_\bfk a_\bfk
			+
			\omega_\bfk b^\dagger_\bfk b_\bfk
		\Bigr)
		+
		\sum_{\bfk n}
			\Bigl(
				A_{\bfk n}^z\rho_\bfk
				+
				A_{\bfk n}^+a^\dagger_{\bar\bfk}+A_{\bfk n}^-a_\bfk
			\Bigr)
			Q_n
	,
\label{eq-Hmagnon}
\end{align}
where $\rho_\bfk=\sum_\bfq a^\dagger_{\bfk+\bfq}a_\bfq$ is the magnon density operator and $\dote{\bfk}=2M(J_0+I_0-J_{\bar\bfk})$ is the magnon energy dispersion in absence of phonons, whereas $\calJ_0=-\lim_{\bfk\rightarrow0}\sum_m\calJ_{mn}e^{i\bfk\cdot(\bfr_m-\bfr_n)}$ with $\calJ=J$ or $I$. Moreover, the magnon-phonon couplings $A_{\bfk n}^z=A_\bfk^ze^{i\bfk\cdot\bfr_n}$ and $A_{\bfk n}^\pm=A_\bfk^\pm e^{i\bfk\cdot\bfr_n}$, where $A_\bfk^z=\sum_mA_{mn}^ze^{i\bfk\cdot(\bfr_m-\bfr_n)}/N$ and $A_\bfk^\pm=-\sqrt{M/2N}\sum_mA_{mn}^\pm e^{\pm i\bfk\cdot(\bfr_m-\bfr_n)}$, respectively, and $\bar{\bfk}=-\bfk$.

\begin{figure}[t]
\begin{center}
\includegraphics[width=\columnwidth]{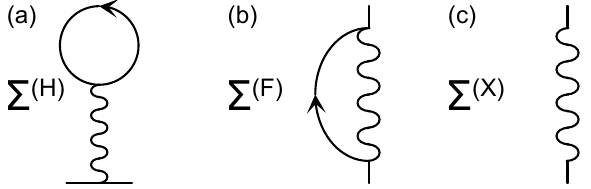}
\end{center}
\caption{Second order self-energy diagrams: (a) Hartree, (b) Fock, or exchange loop, and (c) lowering-raising diagram. Lines with arrows and wiggles represent magnon and phonon propagation, respectively.}
\label{fig-diagrams}
\end{figure}

We consider the magnon properties through the Green function $G_{\bfk\bfk'}(z)=\av{\inner{a_\bfk}{a^\dagger_{\bfk'}}}(z)$. Expanding this propagator order by order in the interaction parameter $A_{\bfk n}$, to lowest non-trivial order we obtain the correction
\begin{align}
G^{(2)}_{\bfk\bfk'}(z)=&
	g_\bfk(z)
		\Bigl(
			\Sigma^{(H)}_{\bfk\bfk'}
			+
			\Sigma^{(F)}_{\bfk\bfk'}(z)
			+
			\Sigma^{(X)}_{\bfk\bfk'}(z)
		\Bigr)
	g_{\bfk'}(z)
	,
\end{align}
where $g_\bfk(s)=1/(z-\dote{\bfk})$ is the unperturbed Green functions, whereas the self-energies are given by
\begin{subequations}
\begin{align}
\Sigma^{(H)}_{\bfk\bfk'}(t)=&
	\sum_{mn\bfkappa}
		A^z_{\bfk-\bfk'm}
		A^z_{\bfkappa m}
		\int
			g^<_{\bfkappa}(\omega)
			\frac{d^>_m(\dote{})-d^<_m(\dote{})}{\dote{}-i\delta}
		\frac{d\omega}{2\pi}
		\frac{d\dote{}}{2\pi}
	,
\\
\Sigma^{(F)}_{\bfk\bfk'}(t,t')=&
	i
	\sum_{m\bfkappa}
		A^z_{\bfk m}A^z_{\bar\bfk' m}
		g_{\bfkappa}(t,t')
		d_m(t,t')
	,
\\
\Sigma^{(X)}_{\bfk\bfk'}(t,t')=&
	2\sum_mA_{\bar\bfk m}^+A_{\bfk'm}^-
	d_m(t,t')
	.
\end{align}
\end{subequations}
These self-energies are depicted in Fig. \ref{fig-diagrams}, showing (a) the Hartree $\Sigma^{(H)}$, (b) the Fock, or exchange loop $\Sigma^{(F)}$, and (c) the lowering-raising $\Sigma^{(X)}$ diagrams. The arrowed lines and wiggles depict the magnon ($g_\bfk$) and phonon ($d_m$) Green function, respectively. The notation used for these self-energies is defined in the following.

\begin{figure}[t]
\begin{center}
\includegraphics[width=\columnwidth]{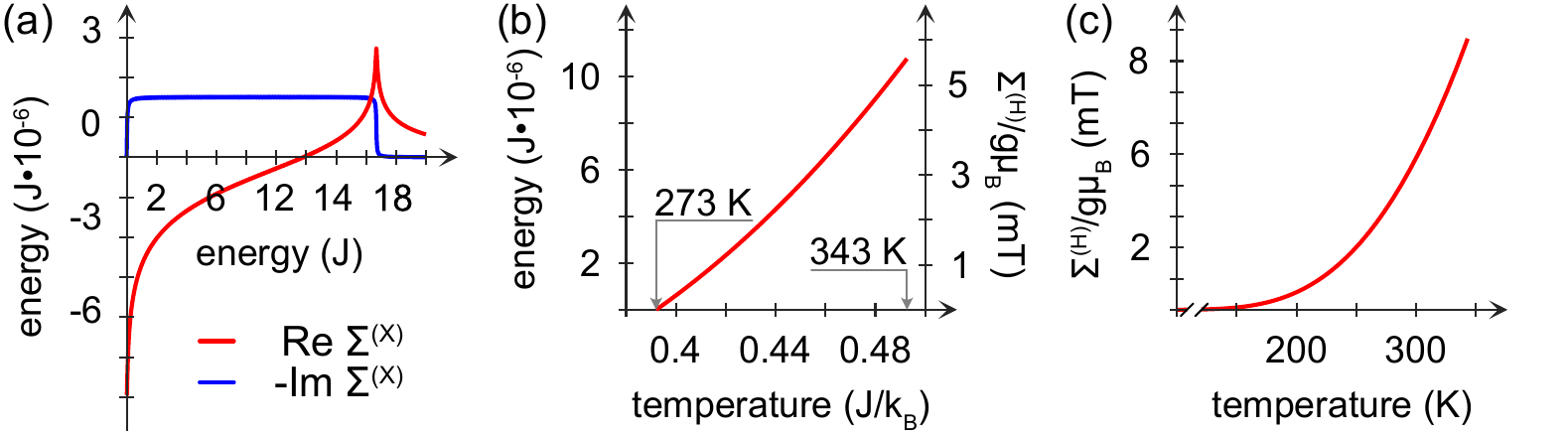}
\end{center}
\caption{Examples of the self-energies (a) $\Sigma^{(X)}$ and (b) $\Sigma^{(H)}$. Here, in (a) $A^\pm_\bfk/J=1/500$ and in (b), (c) $A^z_\bfk=3J\cdot10^{-6}$, and summing over 500 vibrational modes $1/30\leq\omega_m/J\leq50/3$. In (a), the broadening $\Gamma_\text{ph}=0.03J$ has been included for the sake of smoothening the lines. The plots in (b), (c) have been subtraced by $\Sigma^{(X)}(\omega=0)$.}
\label{fig-SigmaX}
\end{figure}

Physically, the diagrams, Fig. \ref{fig-diagrams}, indicate phonon coupling between magnons at different levels. The simplest process is provided in the lowering-raising self-energy $\Sigma^{(X)}$, in which there is an angular momentum exchange between the magnons and phonons, indicated by the raising and lowering transfer rates $A_{\bfk m}^\pm$. To second order, processes of this type contribute an energy shift to the magnetic anisotropy as well as broadening, which indicates dissipation. The plots in Fig. \ref{fig-SigmaX} (a) show an example of the real and imaginary parts of the lowering-raising self-energy. For low energies, this self-energy decreases the magnetic anisotropy while it is increased at high. Simultaneously, the line broadening introduced by the angular momentum transfer via phonons is nearly constant throughout the band width of the phonon spectrum.

The Hartree contribution $\Sigma^{(H)}$ accounts for a simple phonon mediated magnon-magnon interaction, where the lesser magnon Green function $g_\bfq^<$ is proportional to the magnon occupation at the energy $\dote{\bfq}$, whereas the difference between the greater and lesser phonon Green functions, $d^>_m-d^<_m$, accounts for the density of phonon states at the energy $\omega_m$. This self-energy is evaluated by assuming a two-dimensional structure, corresponding to the interface between the ferromagnet and the chiral molecules. Near the $\Gamma$-point, the energy dispersion $J_\bfp\approx J_0+\alpha p^2$, where $\alpha=\sum_mJ_{mn}(\bfr_m-\bfr_n)^2/2$. Making use of that $g_\bfq^<(t,t)=(-i)n_B(\dote{\bfq})$, where $n_B(\omega)$ is the Bose-Einstein distribution function, and $d^>_m(\dote{})-d^<_m(\dote{})=(-i)2\pi[\delta(\dote{}-\omega_m)-\delta(\dote{}+\omega_m)]$, the Hartree contribution becomes
\begin{align}
\Sigma^{(H)}_{\bfk\bfk'}=&
	-\sum_m
		\frac{2A_m^2}{\omega_m}
		e^{i(\bfk-\bfk')\cdot\bfr_m}
		\int
			n_B(\dote{\bfq})e^{i\bfq\cdot\bfr_m}
		\frac{d\bfq}{(2\pi)^2}
\nonumber\\\approx&
	\sum_m
	\frac{A_m^2e^{i(\bfk-\bfk')\cdot\bfr_m}}{4\pi M\omega_m\alpha\beta}
	\ln\frac{1-e^{-2M\beta(I_0-\alpha p_c^2)}}{1-e^{-2M\beta I_0}}
	,
\label{eq-SigmaH}
\end{align}
where $e^{i\bfq\cdot\bfr}$ is replaced by 1, and $p_c$ is an upper cut-off for the quadratic dispersion.For ferromagnetic exchange, $\alpha>0$, this contribution is always positive since $I_0-\alpha p_c^2<I_0$.

With this contribution, the magnon energy is given by
\begin{align}
\dote{\bfk}^{(H)}=&
	\dote{\bfk}
	+
	\sum_m
	\frac{A_m^2}{4\pi M\omega_m\alpha\beta}
	\ln\frac{1-e^{-2M\beta(I_0-\alpha p_c^2)}}{1-e^{-2M\beta I_0}}
	.
\label{eq-dispersions}
\end{align}
Effectively then, the total out-of-plane anisotropy $\tilde{I}_0$ is shifted by the amount represented by the Hartree self-energy. This can be viewed in Fig. \ref{fig-SigmaX} (b), (c), showing a monotonic increase of the additional energy, as function of temperature. In Fig. \ref{fig-SigmaX} (b), the plot is limited to a range between 273 K and 343 K, which is a viable range in a biological context. Here, the anisotropy increases nearly linearly with temperature. The properties of $\Sigma^{(H)}$ for wider temperature range is shown in Fig. \ref{fig-SigmaX} (c). In both these plots, the value $\Sigma^{(X)}(\omega=0)$ has been subtracted.

The exchange loop, $\Sigma^{(F)}$, modifies the magnon spectrum by introducing resonances at the energies $\dote{\bfq}\pm\omega_m$, corresponding to energy exchanges between the magnon and phonon subsystems. On the other hand, this contribution is pinned to the phonon energies and does not add substantially to the magnetic anisotropy, in comparison to the Fock contribution. Therefore, in the following discussion, this contribution will be omitted.

Magnons are excitations from the ferromagnetic ground state of the crystal. Hence, exciting magnons deteriorate the ferromagnetic order which, in turn, is a signature of weakened ferromagnetism. The magnetic anisotropy, on the other hand, defines an energy scale that has be overcome for the magnons to begin to become populated. In other words, the magnetic anisotropy provides stability to the magnetic order that can only be overturned by feeding energy into the crystal. By raising the temperature, magnons may become thermally excited when the energy associated with the temperature is at least as large as the anisotropy. However, if instead, the thermal energy can be channelled into excitations which are not detrimental for the magnetic order, the stability of the magnetic state remains. Here it has been shown that phonons which are coupled to the magnons absorb the thermal energy such that the magnon excitations are suppressed. In addition, the higher the temperature, the more phonons are excited which adds to the magnetic anisotropy such that the magnons become even less available, hence, strengthening the ferromagnetic order.

\subsection*{Entropy production rate}
The purpose is to evaluate the entropy production rate related to the Hartree approximation used above in the discussion of the vibrational contribution to the magnetic anisotropy. The entropy production rate $\dt S$ is provided by the relation $T\dt S=\dt\av{\Hamil}$, for a zero chemical potential. In order to relate the maintained ferromagnetic order to losses, the relevant quantity to calculate is $\dt\sum_\bfk\av{\dote{\bfk}a^\dagger_\bfk a_\bfk}$, since this reflects the gain or loss of entropy due to the coupling to the phonons. Under the assumption that the magnon energy $\dote{\bfk}$ is time-independent, this rate of change is given by
\begin{align}
\dt\sum_\bfk\dote{\bfk}\av{a^\dagger_\bfk a_\bfk}=&
	2
	\sum_{\bfk m}
		\dote{\bfk}
		\im\av{(A^z_{\bfk m}\rho_\bfk+A^+_{\bfk m}a^\dagger_{\bar\bfk})Q_m}
	.
\end{align}

The second contribution, $\im\sum_{\bfk m}\av{A^+_{\bfk m}a^\dagger_{\bar\bfk m}Q_m}$, is conservative to lowest order, and reflects the exchange of energy back and forth between the magnonic and phononic subsystems. By contrast, the first contribution comprises an entropic loss, which to lowest order in the magnon-phonon interactions is given by
\begin{align}
\dt\sum_\bfk&
	\dote{\bfk}\av{a^\dagger_\bfk a_\bfk}
	=
	\Bigl(\dt S\Bigr)_++\Bigl(\dt S\Bigr)_-
	,
\label{eq-dtSlr}
\end{align}
where $(\dt S)_\pm$ denotes the magnonic entropy gain (+) or loss (-) from the magnon-phonon interactions and
\begin{align}
\Bigl(\dt S\Bigr)_\pm=&
	\pm2\pi
	\sum_{\bfk\bfq ms}
		\dote{\bfk}|A_{\bfk m}^z|^2
		n_B(s\dote{\bfq-\bfk})
		n_B(-s\dote{\bfq+\bfk})
\nonumber\\&\times
		n_B(s[\dote{\bfq+\bfk}-\dote{\bfq-\bfk}])
		\delta(\dote{\bfq+\bfk}-\dote{\bfq-\bfk}\pm\omega_m)
		,
\label{eq-dtSpm}
\end{align}
for $s=\pm1$. The physics contained in \eqref{eq-dtSlr} and \eqref{eq-dtSpm} is a manifestation of phonon assisted transitions between different magnon states. The term in this expression with $s=-1$, $n_B(-\dote{\bfq-\bfk})n_B(\dote{\bfq+\bfk})n_B(\dote{\bfq-\bfk}-\dote{\bfq+\bfk})$, reflects the lowering of the magnon energy (and momentum) through the transition $\ket{\bfq+\bfk}\rightarrow\ket{\bfq-\bfk}$ upon emission of the energy quanta $\dote{\bfq-\bfk}-\dote{\bfq+\bfk}$. This energy quanta is subsequently absorbed by the phonon reservoir. The second term ($s=1$) reflects the opposite type of processes. Finally, the presence of the factor with delta functions in \eqref{eq-dtSlr} emphasizes the occurrence of both absorption, $\dote{\bfq+\bfk}-\dote{\bfq-\bfk}-\omega_m$, and emission, $\dote{\bfq+\bfk}-\dote{\bfq-\bfk}+\omega_m$, of magnon energy due to the magnon-phonon interactions. The overall sign of the expression in \eqref{eq-dtSlr} indicates whether the magnons lose ($-$) or gain ($+$) entropy in this connection.

The continuous lowering of the energy of the magnon population by disposing it to the phonon reservoir leads to that the ferromagnetic order is maintained, and with increasing temperature also further stabilized. It may be noticed that $(\dt S)_-<0<(\dt S)_+$, since the summand expressed in \eqref{eq-dtSpm} is positive. Furthermore, by integrating out the angular dependence, the entropy production rate can be written as
\begin{align}
\dt\sum_\bfk\av{a^\dagger_\bfk a_\bfk}=&
	\sum_m
	\int_0^{p_c}
		\frac{\dote{k}|A_{km}^z|^2}{\cos\beta(4MI_0-\alpha[k^2+q^2])+\cosh\beta\omega_m}
\nonumber\\&\times
		\biggl(
			16n_B(\omega_m)
			\sinh\beta(4MI_0-\alpha[k^2+q^2])
\nonumber\\&
			-
			\frac{\omega_m}{M\alpha pk}
			\cosh\beta(4MI_0-\alpha[k^2+q^2])
		\biggr)
	\frac{kdk}{2\pi}
	\frac{qdq}{2\pi}
	.
\end{align}
Both terms in the integrand are negative for a ferromagnet with $I_0<0$. From this observation, we can conclude that the entropy production rate is negative. Hence, the entropy of the magnetic subsystem is lowered by the continuous energy transfer to the phonon subsystem which, in turn, provides further stabilization of the ferromagnetic order.

In conclusion, we have considered ferromagnetism in the presence of chiral phonons. Contrary to conventional perception of magnetic order, we have found that phonons play the role as an energy sink to which thermal energy provided by the environment may be disposed off and, thereby, preventing magnon excitations to become occupied and destroy the ferromagnetic order. A requirement for a viable coupling between the magnetic and vibrational degrees of freedom is (i) absence of inversion symmetry \cite{PhysRevResearch.5.L022039} or (ii) a non-trivial spin-texture within the electronic structure to which the nuclear vibrations are coupled \cite{PhysRevMaterials.1.074404}. The theoretical account provided here explains recent results of increasing coercivity with increasing temperature observed for ribose-aminooxazoline on a ferromagnetic substrate \cite{kapon2024}.

\acknowledgments
J.F. acknowledges support from Stiftelsen Olle Engkvist Byggm\"astare. Y.P. acknowledges the funding from Marie Sklodowska-Curie Actions under Horizon Europe framework program (CISSE project No. 101071886), Carl Zeiss Stiftung (HYMMS project No. P2022-03-044), and U.S Air Force (grant No. FA8655-24-1-7390). Y.K. thanks the Israel Council of Higher Education’s VTT fellowship for women in STEM. S.F.O. and D.D.S. acknowledge the Harvard Origins of Life Initiative for funding and its members for fruitful discussions. S.F.O. also acknowledges the Kavli-Laukien fellowship program and the Kavli and Laukien Foundations for generous research funding and travel support.


\bibliography{pnas-sample}

\end{document}